\documentclass[floatfix,showpacs,epsfig]{revtex4}
\usepackage{pdfpages}
\usepackage{epsf}
\usepackage{subfigure}
\usepackage{amsmath}
\usepackage{epstopdf}
\usepackage{mathrsfs}
\usepackage{natbib}
\usepackage{amssymb,latexsym}
\usepackage{dcolumn}
\usepackage{graphicx}
\def\be{\begin{equation}}
\def\ee{\end{equation}}
\def\ba{\begin{eqnarray}}
\def\ea{\end{eqnarray}}

\graphicspath{{images/}}
\renewcommand{\L}{Lema\^{i}tre}
\begin{document}
\title{\large \bf  Cosmological black holes: the spherical perfect fluid collapse with pressure in a FRW background}
\author{Rahim Moradi}
\affiliation{Department of Physics and ICRA, Sapienza University of Rome,
Aldo Moro Square 5, I-00185 Rome, Italy and\\
ICRANet, Square of Republic 10, I-65122 Pescara, Italy }
\email{Rahim.Moradi@icranet.org}

 \author{Javad T. Firouzjaee}
\affiliation{ School of Astronomy, Institute for Research in Fundamental Sciences (IPM), P. O. Box 19395-5531, Tehran, Iran }
 \email{j.taghizadeh.f@ipm.ir}
\author{Reza Mansouri}
\affiliation{Department of Physics, Sharif University of Technology,
Tehran, Iran and \\
  School of Astronomy, Institute for Research in Fundamental Sciences (IPM), Tehran, Iran}
 \email{mansouri@ipm.ir}

\begin{abstract}
We have constructed a spherically symmetric structure model in a cosmological background filled with perfect fluid with non-vanishing pressure as an exact solution of Einstein equations using the \L\ solution. To study its local and quasi-local characteristics including the novel features of its central black hole, we have suggested an algorithm to integrate the equations numerically. The result shows intriguing effects of the pressure inside the structure. The evolution of the central black hole within the FRW universe, its
decoupling from the expanding parts of the model, the structure of its space-like apparent horizon, the limiting case of the dynamical horizon tending to a slowly evolving horizon, and the decreasing mass in-fall to the black hole is also studied. The cosmological redshift of a light emitted from the cosmological structure to an observer in the FRW background is also calculated. This cosmological redshift includes local and cosmic part which are explicitly separated. We have also formulated a modified NFW density profile for a structure to match the exact solution conditions.

\end{abstract}
\maketitle
\section{introduction}

The term cosmological black hole (CBH) is used to describe a collapsing structure within an otherwise expanding universe after the
radiation era. This is different from the term astrophysical black holes (ABHs) coined to use for the applications of static or stationary
asymptotically flat black holes in astrophysics (see for example recent paper \cite{ellis-13}). CBHs are
fundamentally different from the ABHs in being non-stationary and having a dynamical horizon, although being as
much interesting as ABHs for the astrophysical applications and well suited to be used as a test arena for fundamental physics
in strong gravity and quasi-local phenomena in weak gravity regimes. Despite extensive studies in relativistic structure
formation and static and stationary black holes we still have little information based on exact general relativistic
studies regarding the main features of such overdense cosmological regions including its central black hole. \\
Since the early beginning of the discovery of the expansion of the universe people have been looking for models
describing an overdense region in a cosmological background (\cite{McVittie}, see also \cite{cosmological black hole}).
In the matter dominated era the initial expansion of an overdense region within an expanding universe will finally
decouple from the background and collapses to a dynamical black hole, in contrast to the overdensity regions in the
radiation dominated era in the early universe where the density perturbation has to be of the order of the horizon
to collapse to a primordial black hole (PBH). In the matter dominated era, we expect a CBH within the resulting structure and
a very weak gravitational field outside it, being different from the familiar Schwarzschild one in being neither
static nor asymptotically flat \cite{man}. Therefore, such cosmological structures, if based on exact solutions of
general relativity and not produced by a cut-and-paste technology, are very interesting laboratories to study not only general
relativistic non-linear structures, their quasi-local features such as mass and horizons \cite{man}, black holes thermodynamics
and information loss puzzle \cite{giddings-12, manradiation}, but the validity of the weak field approximation as well \cite{mojahed}. After all, the
universe is evolving and asymptotically not Minkowskian. Therefore, one needs to have a dynamical model for a black hole, to
be compared with the familiar results in the literature on black holes within a static and asymptotically flat space-time \cite{waldbook},
where global concepts such as the event horizon are not defined. Nevertheless, depending on the specific cosmological model we may be able
	to define a hypersurface separating the trapped null outgoing geodesics from those approaching (future-)infinity. This hypersurface effectively plays
    the role of the event horizon \cite{man}. Numerically, this hypersurface is determined by tracing the outgoing null geodesics at late times; We will refer 
    to it as the quasi event horizon. \\

The need for a local definition of black holes and their horizons has led us to concepts such as Hayward's trapping
horizon {\cite{Hayward94}, isolated horizon \cite{ashtekar99}, Ashtekar and Krishnan's dynamical
horizon (DH) \cite{ashtekar02}, and Booth and Fairhurst's slowly evolving horizon {\cite{booth04}. The CBH we are going
to study is an excellent example of testing these different concepts and their relationship in addition to understand its
difference to an asymptotically Minkowskian and static black hole. Now, a widely used metric to describe the gravitational
collapse of a spherically symmetric dust cloud is the so-called Tolman-Bondi-\L (LTB) metric \cite{LTB}. Exact general
relativistic models for the dynamic of an asymptotically FRW structure leading to a central dynamical black hole based on
a LTB metric with no pressure has been reported in \cite{man, mangeneral}. These models may be extended to a perfect fluid
with a non zero pressure, the so-called \L \ models \cite{hellaby}. Our interest is now using these inhomogeneous cosmological
solutions of Einstein equations as a model for a cosmic structure with non-vanishing internal pressure leading to a dynamical
CBH within a FRW matter dominated expanding universe. We, therefore, try an ideal fluid with an interpolating pressure function
being non-zero inside the structure and vanishing at infinity where we expect a matter dominated FRW universe. To achieve
this goal, we have to avoid any cut-and-paste technology of finding the solution. Any internal solution pasted to a FRW universe
does not reflect the dynamics of the inhomogeneous universe due to the homogeneity of FRW universe outside the structure. In
contrast, our model is just asymptotically FRW reflecting the full relativistic local and quasi-local effects due to the cosmic
fluid. The ideal fluid is modeled such that the non-vanishing pressure inside the structure goes smoothly to a matter dominated
universe far from the structure. A similar problem studied extensively in literature in the last 40 years is the issue of
primordial black holes (PBHs). These are structures within the radiation dominated phase of the universe, usually in the
late phase of inflation, due to the superhorizon perturbations \cite{pbh}. PBHs are usually based on the same \L\ cosmological
solutions of Einstein equations which are spherically symmetric and inhomogeneous. The perturbation is formalized by
assuming that the horizon size $R_H$ is smaller than the structure size $R_S$, or by assuming
$\epsilon \equiv \frac{R_H}{R_S}< 1$, where the relevant quantities are expanded in the powers of $\epsilon$.
Using the same \L\ solution although not in the same coordinates, we are interested in cases where
$\epsilon \gg 1$, i.e. structures are much smaller than the horizon size as expected for cosmological structures
within the matter dominated FRW universe. Therefore, we expect to recover novel features and we have also to propose a new algorithm how to solve numerically
the field equations. In addition we are asking different questions such as the kind of black
holes we may encounter and the characteristics of their apparent and event horizons, the effect of the pressure
inside the structure on the collapse behavior and the matter flux. There are many other questions encountering
in black hole literature to be faced in future studies, such as the definition of a non-rotating spherically
symmetric dynamical black hole within an expanding universe, its differences to the simple Schwarzschild or Kerr
black hole, its event and apparent horizon and their features, the internal structure of CBH, the rate of collapse,
the effect of pressure inside the structure and its probable non-Newtonian and non-linear novel effects due to
the non-vanishing matter outside, and the information puzzle. That is why the study of CBHs goes beyond the
study of PBHs and is a new arena for a novel black hole terminology.\\
Section II is an introduction to the spherically symmetric inhomogeneous prefect fluid cosmological models. In section III the result of the numerical integration is reported expressing the main characteristics of our
model assuming different pressure profiles. In the section IV, the redshift of the light coming from nearby the structure is investigated.  In section V we develop a modified NFW density profile for a cluster of galaxies taking into account the requirements of our solution and study its evolution. We will then discuss the result in section VI. Throughout the paper we assume $8\pi G = c = 1$.\\

\section{General spherically symmetric solution}

   Consider a general inhomogeneous spherically symmetric spacetime \cite{hellaby} filled with a perfect fluid and a metric expressed in the comoving coordinates, $x^{\mu} = (t,\,r,\,\theta,\,\phi)$:
\begin{equation}\label{metric}
ds^2  = -e^{2\sigma} \, dt^2  + e^{\lambda} \, dr^2 + R^2 \, d\Omega^2 \;,
\end{equation}
 where $\sigma = \sigma(t,r)$, $\lambda = \lambda(t,r)$ are functions to be determined, $R = R(t,r)$ is the physical radius, and $d\Omega^2 = d\theta^2 + \sin^2 \theta \, d\phi^2$ is the metric of the unit 2-sphere.  The energy momentum tensor of the perfect fluid is given by
 \begin{equation}\label{mT}
   T^{\mu\nu} = (\rho + p) \, u^{\mu} \, u^{\nu} + g^{\mu\nu}p \;,
 \end{equation}
 where $\rho = \rho(t,r)$ is the mass-energy density, $p = p(t,r)$ is the  pressure,  and $u^{\mu} = (e^{-\sigma}, 0, 0, 0)$ is the perfect fluid four-velocity.

\subsection{Field Equations} \label{FEq}

  In addition to the Einstein field equations, $G^{\mu\nu} = \kappa \,T^{\mu\nu} - g^{\mu\nu}\Lambda$, we will use the
   conservation equations in the form \newline
    \begin{equation}
   \frac{2 e^{2\sigma}}{(\rho + p)} \, \nabla_\mu T^{t\mu}  = \dot{\lambda}
      + \frac{2 \dot{\rho}}{(\rho + p)} + \frac{4 \dot{R}}{R} \; = 0
      \label{gtt}
       \end{equation}
   \begin{equation}
   \frac{e^{\lambda}}{(\rho + p)} \, \nabla_\mu T^{r\mu}  = \sigma' + \frac{p'}{p + \rho} = 0,
      \label{grr}
 \end{equation}
where the dot means the derivative with respect to $t$, and the prime means the derivative with respect to $r$.
The Einstein equations lead finally to following equations
\begin{equation}\label{ev}
   \frac{\partial}{\partial r} \left[ R + R \dot{R}^2 e^{-2\sigma} - R  R'^2 e^{-\lambda} - \frac{1}{3} \Lambda R^3 \right] = \kappa \rho R^2 R', \;
 \end{equation}
  and
\begin{equation}\label{Pr}
   \frac{\partial}{\partial t}\left[ R + R \dot{R}^2 e^{-2\sigma}-RR'^2 e^{-\lambda}-\frac{1}{3}\Lambda R^3\right] = -\kappa p R^2\dot{ R}. \;
   \end{equation}
 The term in the brackets is related to the Misner-Sharp mass, $M,$ defined by
 \begin{equation}
   \frac{2M}{R} = \dot{R}^2 e^{-2\sigma} - R'^2 e^{-\lambda} + 1 - \frac{1}{3} \Lambda R^2 \;.
   \label{mR}
 \end{equation}
 Eqs (\ref{ev}) and (\ref{Pr}) may now be written as
 \begin{equation}
   \kappa \rho  = \frac{2M'}{R^2 R'} ~,
   \kappa  p  = -\frac{2\dot{M}}{R^2 \dot{R}} ~.   \label{pressure}
 \end{equation}
 We may write Eq (\ref{mR}) in the form of an evolution equation of the model:
 \begin{equation}
   \label{Ev}
      \dot{R} = \pm e^{\sigma} \sqrt{\frac{2M}{R} + f + \frac{\Lambda R^2}{3}} ~, \\
 \end{equation}
where
\begin{equation}
   \label{f}
      f(t, r) = R'^2 e^{-\lambda} - 1~
 \end{equation}
 is the curvature term, or twice the total energy of test particle at $r$ (analogous to $f(r)$ in the LTB model). In this paper we have set $\Lambda$=0. Note that $R(t,r)$ can not be directly obtained from this
 equation because of the unknown functions $\lambda$, $\sigma$, and $M$.

   The metric functions $g_{tt}$ and $g_{rr}$ may be obtained by integrating (\ref{grr}) and (\ref{gtt}):
 \begin{equation}
   \label{si}
      \sigma  = \ c(t) - \int_{r_0}^r  \frac{ p' \, \ dr}{(\rho + p)}|_{t=const}
      = \sigma_0 -\int_{\rho_0}^\rho \frac{(\frac{\partial p}{\partial\rho})}{(\rho + p(\rho))} \, d\rho~|_{t=const},
 \end{equation}
 and
 \begin{equation}
   \label{lamb}
      \lambda  = \lambda_0(r) - 2 \int_{\rho_0}^\rho
      \frac{d\rho}{(\rho + p(\rho))}
      - 4 \ln \left( \frac{R}{R_0} \right)~|_{t=const},
 \end{equation}
 where $c(t)$ and $\lambda_{0}(r)$ are arbitrary functions of integration (see \cite{hellaby} for more details). In the case of $c(t)$ it is easily seen that requiring our coordinates to lead
 to the LTB synchronous ones for $p = 0$ leads to $c(t)=0$. We notice also that according to (\ref{f}) and the LTB coordinate conditions, the choice of
 $\lambda_{0}(r)$ is equivalent to the choice of $f(t_0,r)=f_0(r)$. One may  prefer to choose $f_0(r)$ and then calculate $\lambda_0(r)$
 from $e^{\lambda_0} = R_0'^2/(1 + f_0)$.\\
  We have therefore 5 unknowns $p, \rho, \sigma, \lambda$, and $R$, four dynamical equations $\dot{\rho}$, $\dot{M}$, $\dot{\lambda}$,
and $\dot{R},$ in addition to an equation of state $p = p(\rho)$, and the definition of the mass $M$ (\ref{mR}).
This defines a numerical algorithm to find solutions for the dynamics of our spherical structure after assuming the initial conditions. The algorithm for solving the coupled PDEs can be found in \cite{mine}

\section{equation of state and the results}

We are now ready to specify the equation of state and integrate the model to see its characteristics. To have a comparative discussion of the
results we consider two types of equation of state: a perfect fluid with a constant state function, $p=w\rho$, and a more general case with
the equation of state $p=w s(r)\rho$ matching our needs for a structure with pressure inside the structure and a pressure-less matter dominated universe far from
the structure. We may then choose the function $ s(r)$ in a way that the pressure becomes zero at infinity, i.e. at  $r>>r_0$. A suitable choice is
$s(r)=e^{-\frac{r}{r_0}}$ where $r_0$ is the distance of the void (boundary of expanding and collapsing phase) from the center of the structure. This is a more realistic model to describe a black hole collapse within the FRW universe and to see the effect of the inside
pressure while the universe outside is matter dominated with no pressure.\newline{}

The model we envisage starts from a small inhomogeneity within a FRW universe. The density profile should be such that the metric outside the structure
tends to FRW independent of the time while the central overdensity region undergoes a collapse after some initial expansion. At the initial conditions, where the
density contrast of the overdensity region is still too small, we may assume that the metric is almost FRW or LTB; the density contrast and the
pressure does not play a significant role. The dynamics of \L\ universe will give us anyhow the expected structure at late times. To choose the initial
conditions at the time $t_0$, we will therefore use a LTB solution with a negative curvature function. We have in fact tried both examples of LTB
or FRW initial data and received no significant difference between the final \L\ solutions. \\
Now, let us choose the the two initial functions $f(r)=f(t_0,r)$ and $M(r)=M(t_0,r)$ in the following way to achieve an asymptotically FRW final solution:

\begin{equation}
f(r)=-\frac{1}{b}re^{-r},
\end{equation}
\begin{equation}
M(r)=\frac{1}{a}r^{3/2}(1+r^{3/2}).
\end{equation}
Far from the central overdensity region we have

\begin{equation}
lim_{r\rightarrow\infty}f(r)=0,
\end{equation}
\begin{equation}
lim_{r\rightarrow\infty}M(r)=\frac{r^3}{a},
\end{equation}
showing the asymptotically FRW behavior of the initial conditions. The corresponding LTB solution of Einstein equations now gives us $R(r)=R_0(r)$ at
the initial time $t_0$. By choosing suitable a and b the "reality condition"\cite{Joshi}
\begin{equation}
\frac{2M}{R} + f + \frac{\Lambda R^2}{3}>0,
\end{equation}
 will be satisfied. This condition can be obtained from Eq (\ref{Ev}). In addition to "reality condition", the "weak energy condition" \cite{Joshi} must be satisfied at every r and t:
 \begin{equation}
\rho(r,t)\geq 0, \quad \rho(r,t)+p(r,t)\geq 0,
\end{equation}
  It can be seen from Fig.(\ref{den2}) that our initial conditions give rise to $\rho(r,t)\geq 0 $ at every t and r.\newline{}

 Assuming an equation of state to satisfy weak energy condition, $p(r,t)\geq 0 $, is now enough to numerically calculate the necessary dynamical functions of the model.
Specifically, by looking at $\dot{R}(t,r)$ and $\rho(t,r)$ we may extract informations of how the central region starts collapsing after the initial
expansion and how a black hole with distinct apparent and event horizons develops while the outer region expands as a familiar FRW universe. We may also
find out the difference to the case of the pressure-less model. It will also show if and how the very weak gravity outside the collapsed structure
affects the dynamic of the central structure in comparison to the familiar Schwarzschild model. The results of the numerical calculation for both
equation of states are given in the following sections.\\

\subsection{The density behavior}

 The density profiles for both equation of states as a function or $t$ and $r$ are given in Fig.(\ref{den2}). A comparison of
these figures shows the effect of the pressure on the development of the central black hole. Obviously in case of non-vanishing pressure outside
the structure the collapse is more highlighted with a more steep density profile. The over-density region in the collapsing phase is always
separated from the expanding under-density region through a void not expressible in these figures. We will consider the deepest place of the
void as the boundary of the structure. This boundary is always near by the boundary of the contracting and the expanding region of the
model structure.\newline{}
\begin{figure}
  \centering
  \mbox{
    \subfigure[Density evolution of our cosmological black hole for the perfect fluid with the equation of state $p=w\rho$.\label{den1}]{\includegraphics[width=0.5\linewidth]{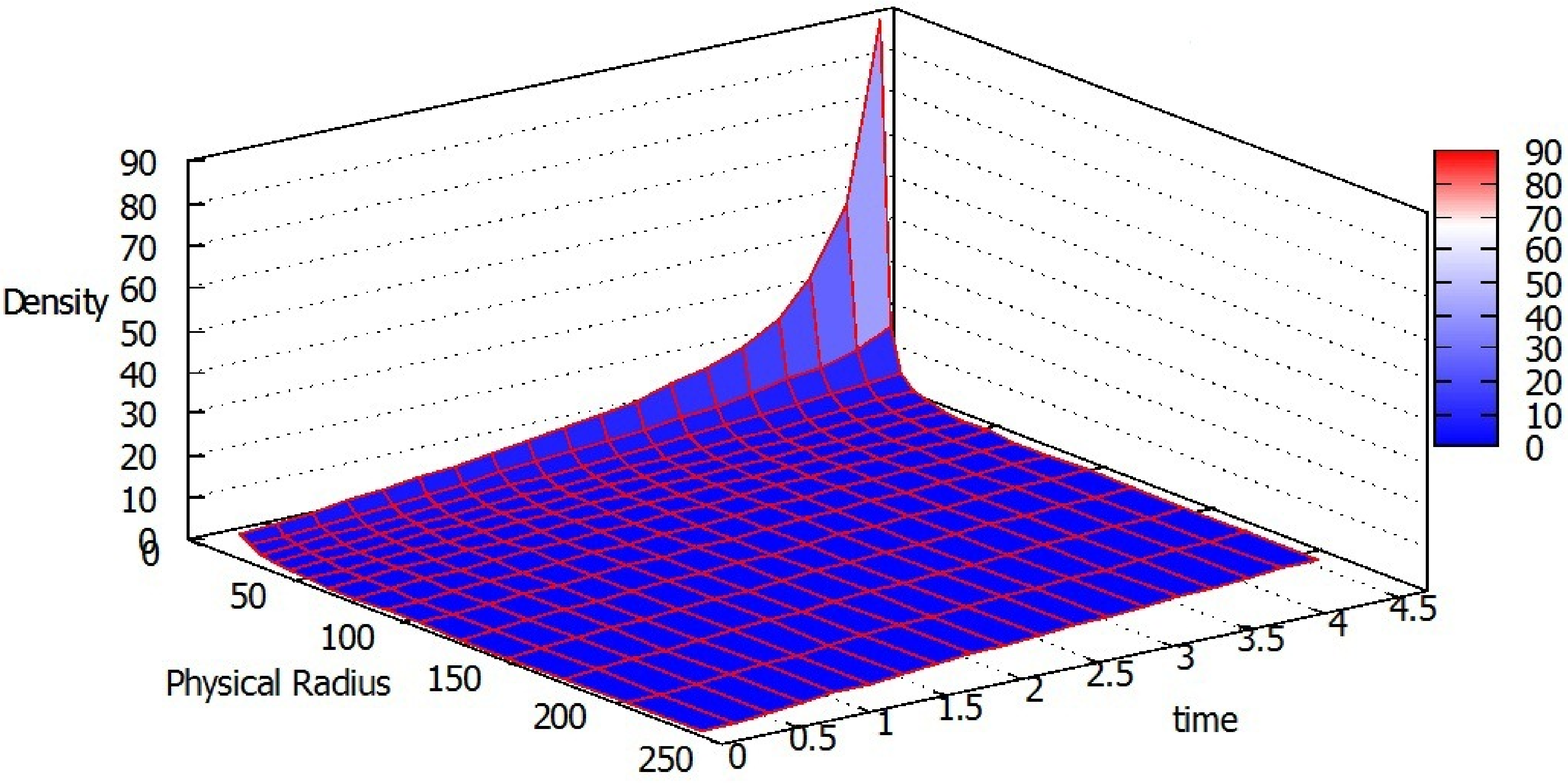}} \quad
   \subfigure[Density evolution of our cosmological black hole for the perfect fluid with the equation of state $p=w\rho s(r)$. Note the less significant central density and the more flat density profile near the center of the structure.\label{1kol}]{\includegraphics[width=0.5\linewidth]{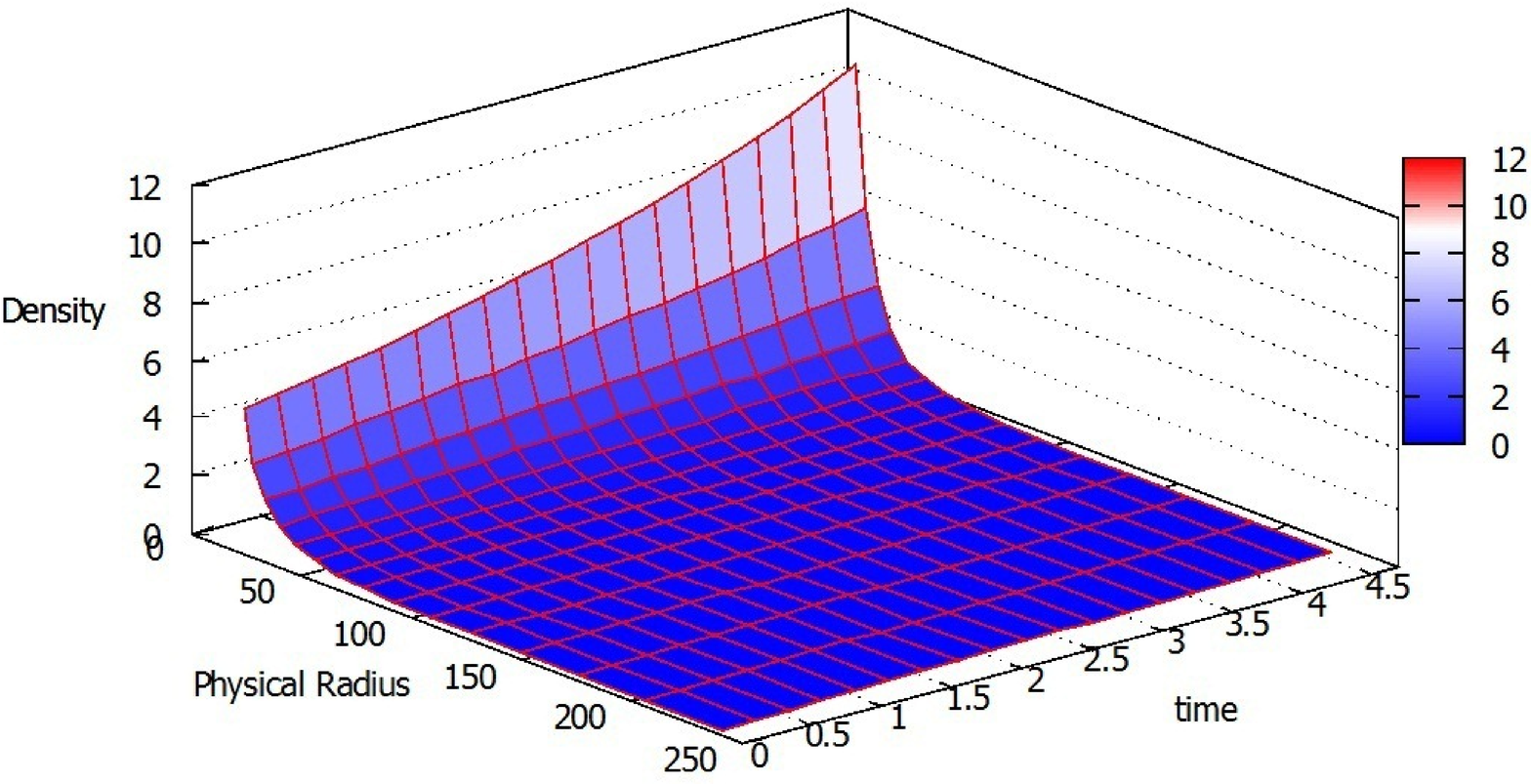}}
  }
  \caption{Density evolution of our cosmological black hole.}
  \label{den2}
\end{figure}

 In the numerical calculation at late times the density near the center of the black hole approaches infinity leading to numerical errors and 
	crashing of the algorithm. We refer to this effect as the black hole singularity, as discussed in \cite{man, mine}. The cosmological singularities (big bang and big crunch) are not included in our numerical simulation due to the initial conditions being fixed at a time much later than the big bang in the matter dominated era. Assuming the background 
	to be the standard flat FLRW metric, there is no big crunch singularity in our model. By selecting other backgrounds such as the closed FLRW model, the big crunch singularity
	will play a crucial role in the development of singularities \cite{man}.

\subsection{ The pressure effect}

Figs.(\ref{2kol}) and (\ref{22kol}) show the behavior of the collapsing and the expanding regions for the equation of state $p = w\rho$ by
depicting the corresponding \L\ Hubble parameter $\dot R/R$ versus the physical radius $R$. Figs. (\ref{1kol}) and (\ref{11kol}) show
similar data for the equation of state $p = ws(r)\rho$. Note that the function $s(r)$, as defined to get matter dominated FRW universe
at far distances, has no significant effect, and the qualitative behavior of the dynamics of the physical radius is independent of it.
Therefore, as far as we are interested in the qualitative features of the model, we will just use the simple equation of state with
$s(r)=1$.\\
The place of separation between the expanding and collapsing region defined by $\dot{R}>0$ and $\dot{R}<0$ is almost coincident with
the place of the void where we have defined as the boundary of the structure. Now, from the  figures we realize that the effect of
the pressure in different regions of the model and its comparison to the homogeneous FRW model is an intriguing one. As we know already
from the Friedman equations in FRW models, the pressure adds up to the density and has an attractive effect slowing down the expansion
leading to a more negative acceleration ($\frac{\ddot{a}}{a}=-\frac{1}{6}(\rho+3 p)$). This is evident from the figures at distances far
from the center where our model tends to an FRW one. Whereas within the structure where we have a contracting overdensity region the
behavior is counter-intuitive. Except for the case of vanishing pressure, in all the other cases the pressure begins somewhere to act
classically like a repulsive force opposing the collapse of the structure. To see this more clearly, we have also depicted the acceleration
in Fig.(\ref{accelaration}). As we approach distances near to the center, the negative acceleration in the FRW limit and even inside the
void, gradually increases to positive values, meaning that somewhere within the structure the contraction of the structure
slows down due to the pressure like a classical fluid. Therefore, the pressure effect begins somewhere within the structure to act like a
repulsive force in contrast to the outer regions where its attractive nature dominates. Note that the central black hole and its horizon has
a much smaller radius than the region of the repulsive pressure effect we are discussing.

\begin{figure}
  \centering
  \mbox{
    \subfigure[The overall scheme: $\dot{R}>0$ and $\dot{R}<0$ show the expanding and collapsing regions, respectively.\label{2kol}]{\includegraphics[width=0.5\linewidth]{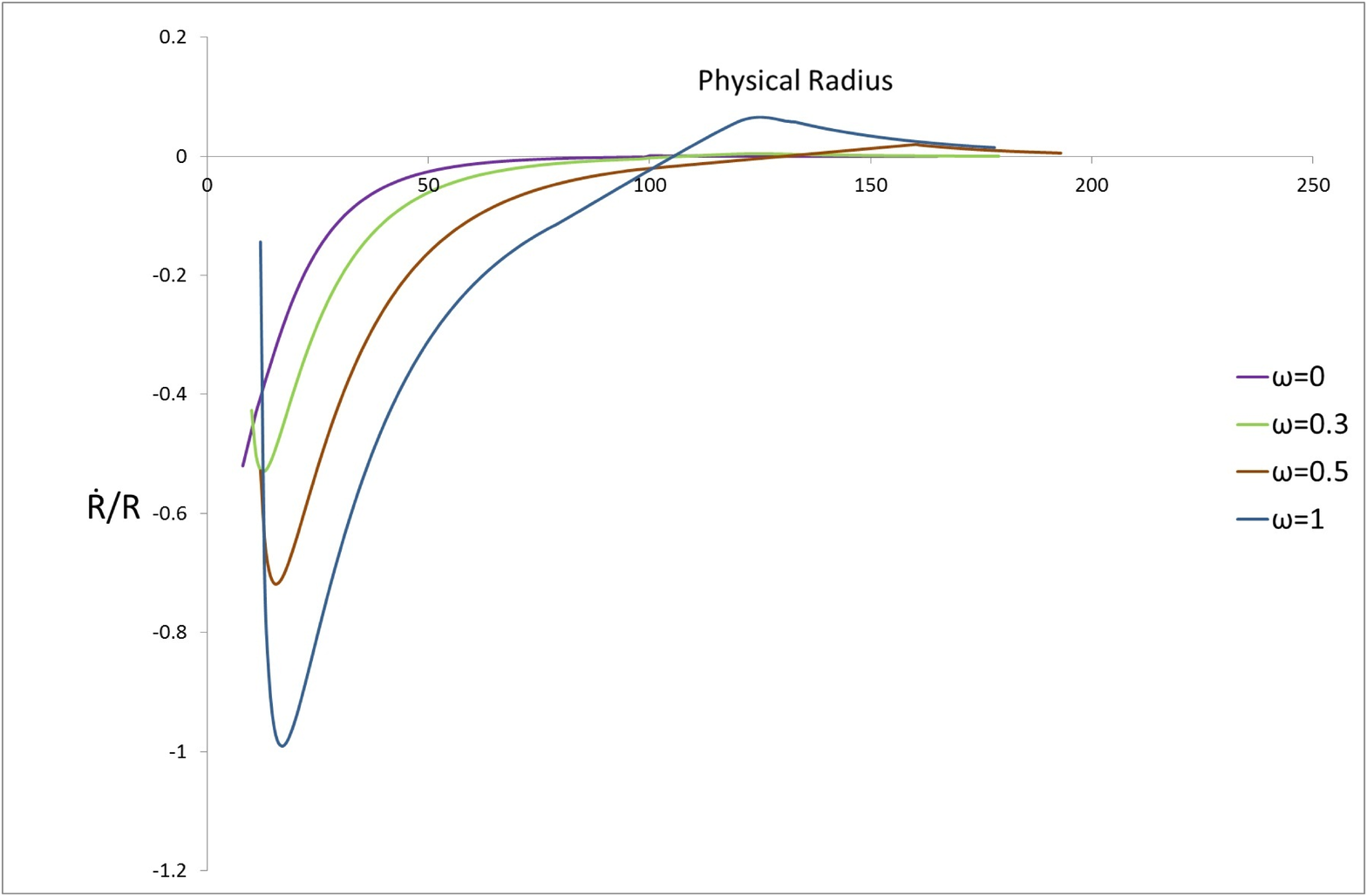}} \quad
   \subfigure[A magnified scheme related to the regions near to the center\label{1kol}]{\includegraphics[width=0.5\linewidth]{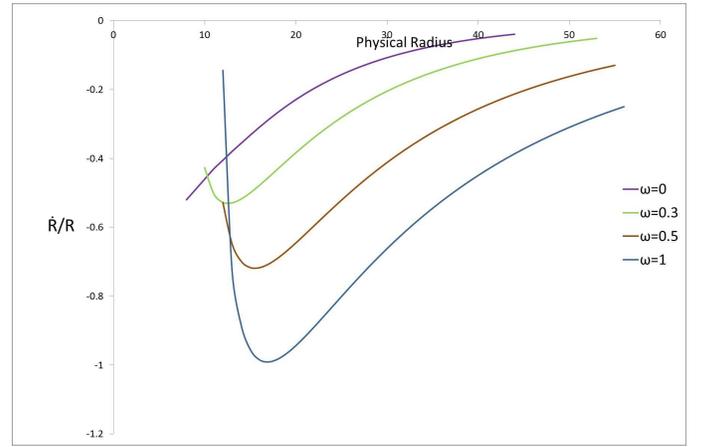}}
  }
  \caption{The behavior of the Hubble parameter $\dot{R}/R$  in the case of $p=w\rho$. Evidently the pressure slows down the
collapse velocity near the center of the structure}
  \label{main figure label}
\end{figure}

\begin{figure}
  \centering
  \mbox{
   \subfigure[The overall scheme: $\dot{R}>0$ and $\dot{R}<0$ show the expanding and collapsing regions, respectively. \label{22kol}]{\includegraphics[width=0.5\linewidth]{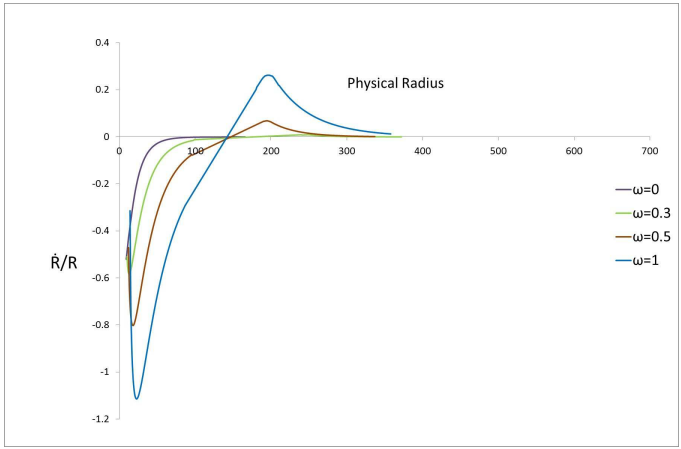}}\quad
    \subfigure[A magnified scheme related to the regions near to the center.\label{11kol}]{\includegraphics[width=0.5\linewidth]{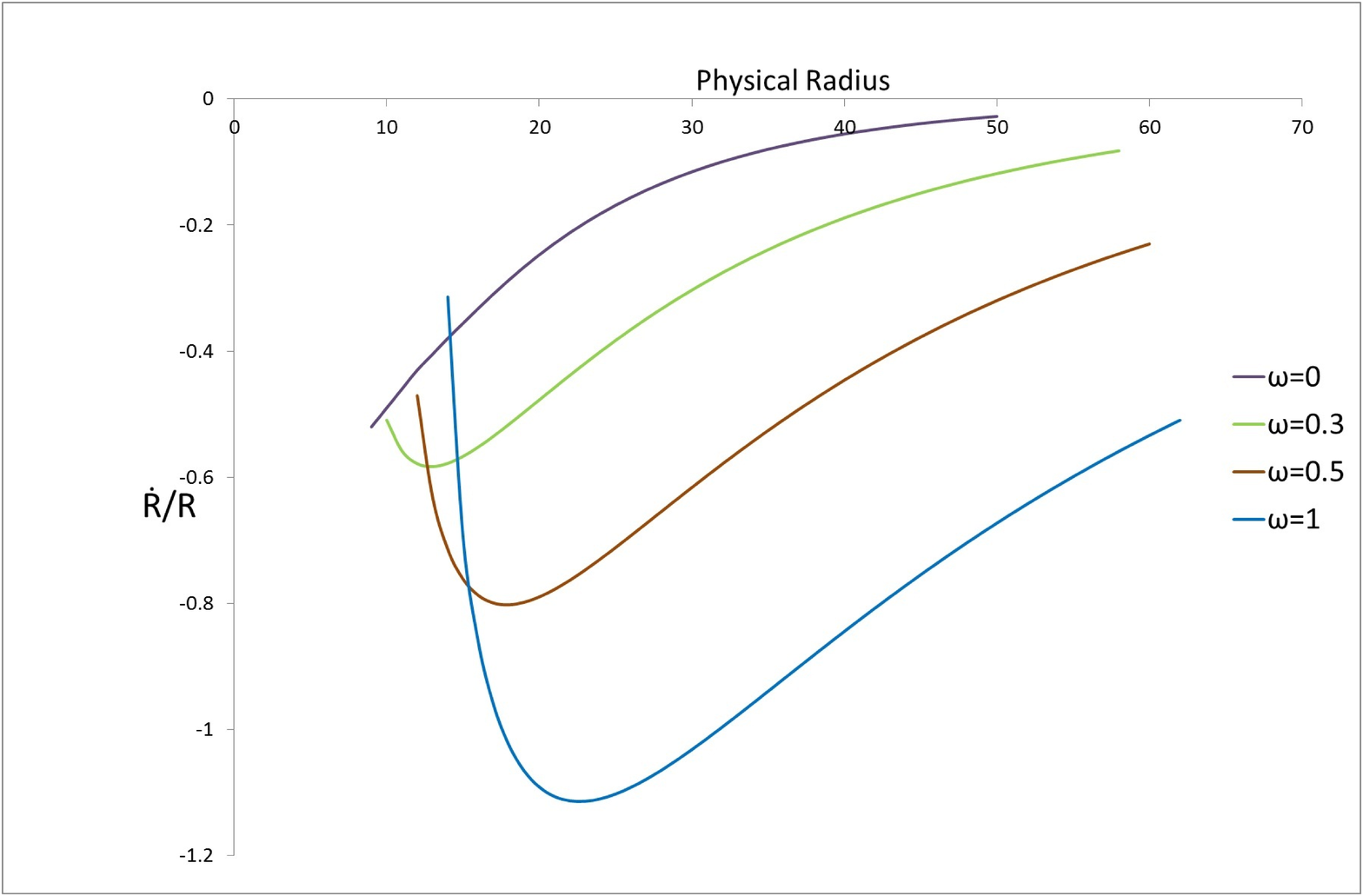}}
  }
  \caption{The behavior of the Hubble parameter $\dot{R}/R$ in the case of $p=w\rho s(r)$. The features are qualitatively as in Fig (\ref {main figure label}).}
  \label{main figure label1}
\end{figure}

\begin{figure}
  \centering
  \mbox{
   \subfigure[The overall scheme: $\dot{R}>0$ and $\dot{R}<0$ show the expanding and collapsing regions, respectively. \label{acc}]{\includegraphics[width=0.5\linewidth]{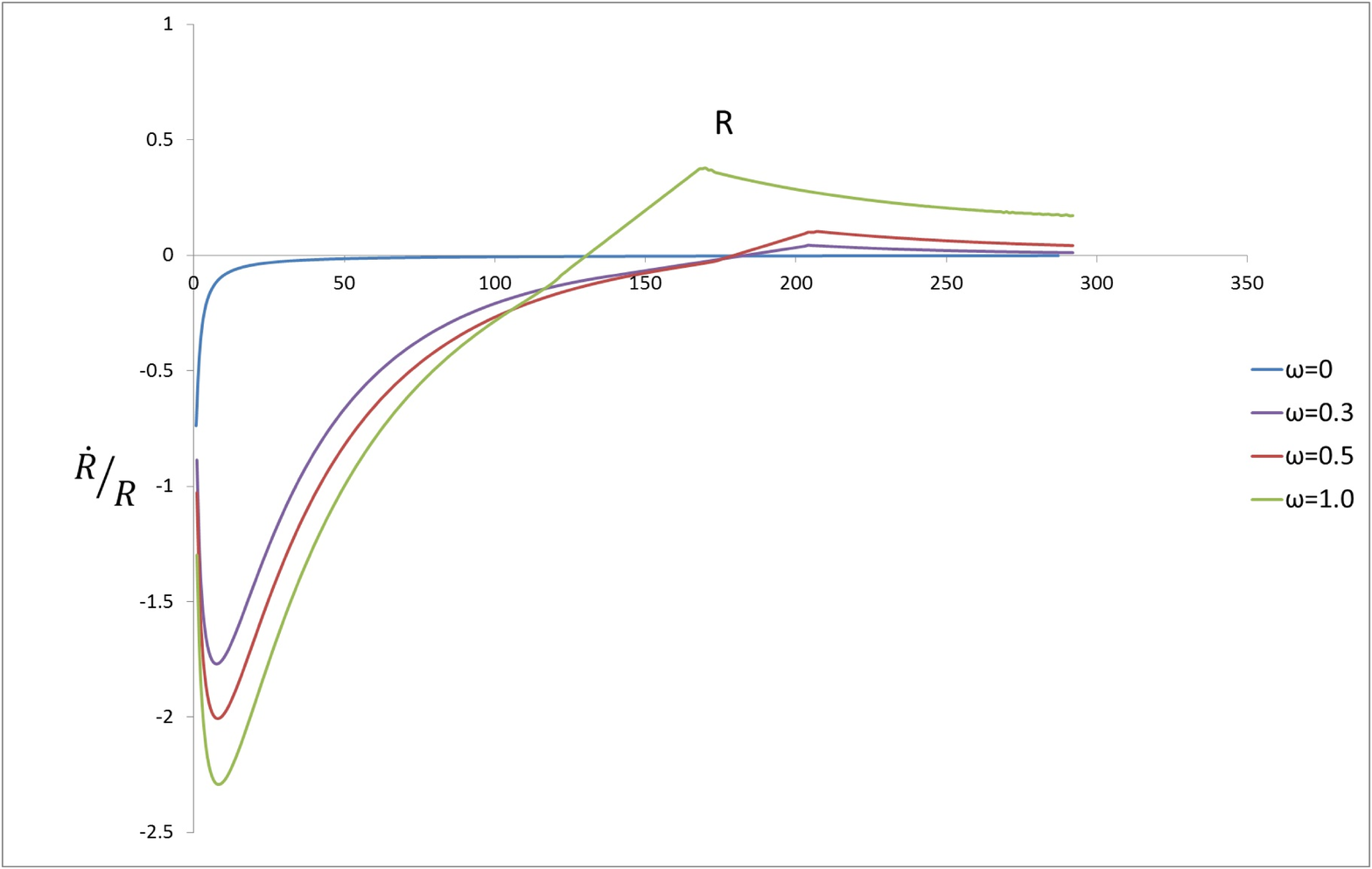}}\quad
    \subfigure[The overall scheme of the acceleration.\label{acc1}]{\includegraphics[width=0.5\linewidth]{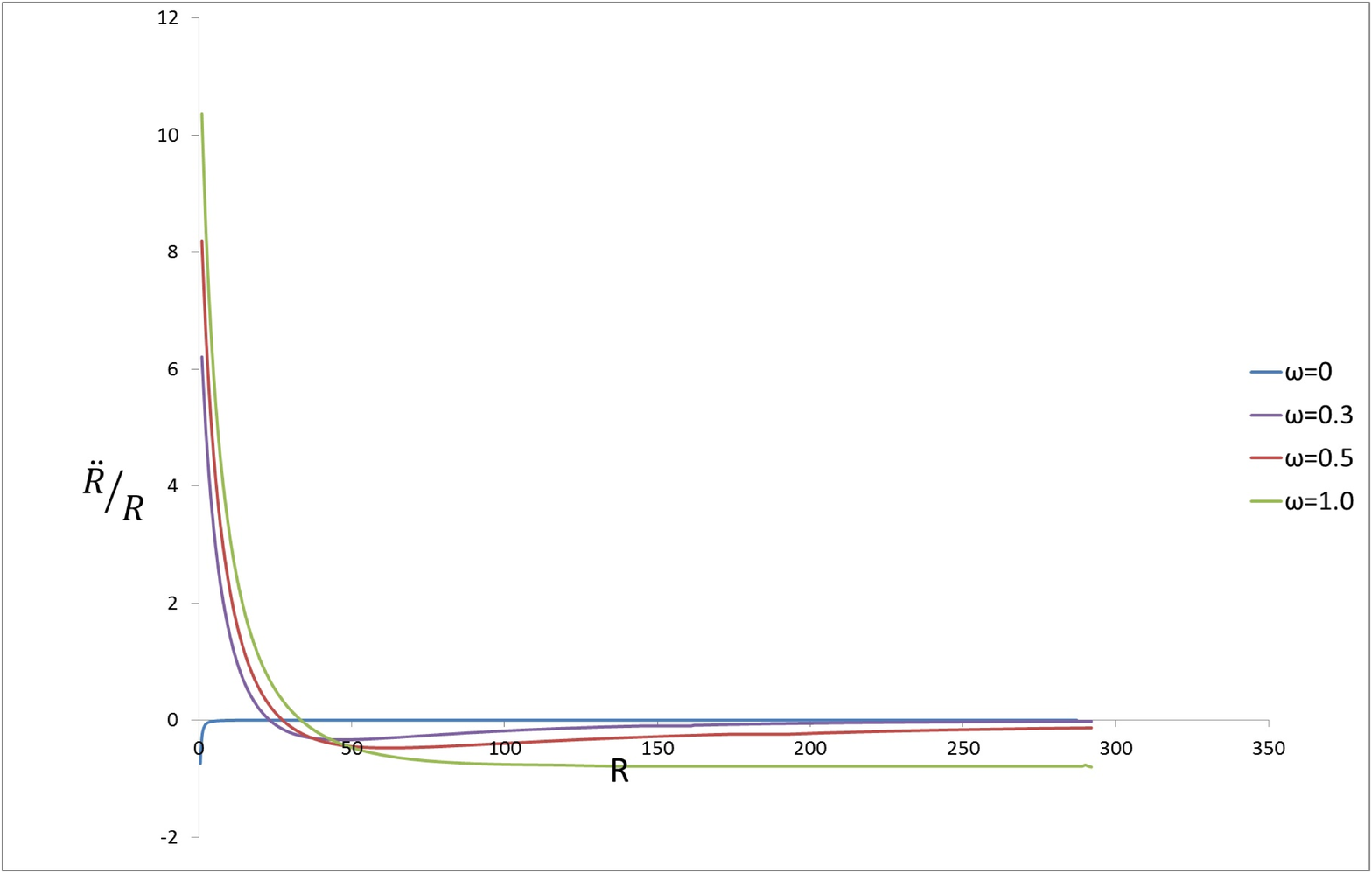}}
  }
  \caption{The behavior of the Hubble parameter $\dot{R}/R$ and the acceleration $\ddot{R}/R $ in the case of $p=w\rho $. }
  \label{accelaration}
\end{figure}

\subsection{The Apparent and Quasi Event Horizon}

The boundary of a dynamical black hole, where the area law and the black hole temperature are defined, is a non-trivial concept (see for
example \cite{man}, \cite{Faraoni} and \cite{manradiation}). Our model is again a concrete example to see the behavior of both apparent and quasi event horizon of
a dynamical structure within an expanding universe. It is easily seen that the apparent horizon of our cosmological black hole is located
at $R=2M$ \cite{mangeneral}. We note that after the formation of an apparent horizon the central density goes to infinity indicating the formation of 
	the singularity at the center. Therefore, the singularity is covered by the dynamical horizon and is not naked!

This apparent horizon is calculated in $t, r$ coordinates numerically.  It is always space-like tending to be light-like at late times. This can best be seen by comparing the slope of the apparent horizon relative to the light cone at every coordinate
point of it. This is in contrast to the Schwarzschild black hole horizon where it is always light-like. At the late times, however, we expect
the apparent horizon to become approximately light-like and approaching the event horizon. This is reflected in
the Fig.(\ref{horizon3}). It is evident that $ \frac{dt}{dr}|_{AH}< \frac{dt}{dr}|_{null}$ at all times on the apparent horizon, the difference tending to zero at
late times. Therefore, the apparent horizon is always a space-like \emph{dynamical horizon} leading to a \emph{slowly varying horizon} at late times \cite{ashtekar02, man}. Note
that the qualitative result is independent of the equation of state.
\begin{figure}
  \centering
  \mbox{
   \subfigure[The $p=w\rho$  case: $ \frac{dt}{dr}|_{AH}< \frac{dt}{dr}|_{null}$  on the apparent horizon. Therefore, the
apparent horizon is always a space-like dynamical horizon leading to a slowly varying horizon at late times. \label{horizon1}]{\includegraphics[width=0.5\linewidth]{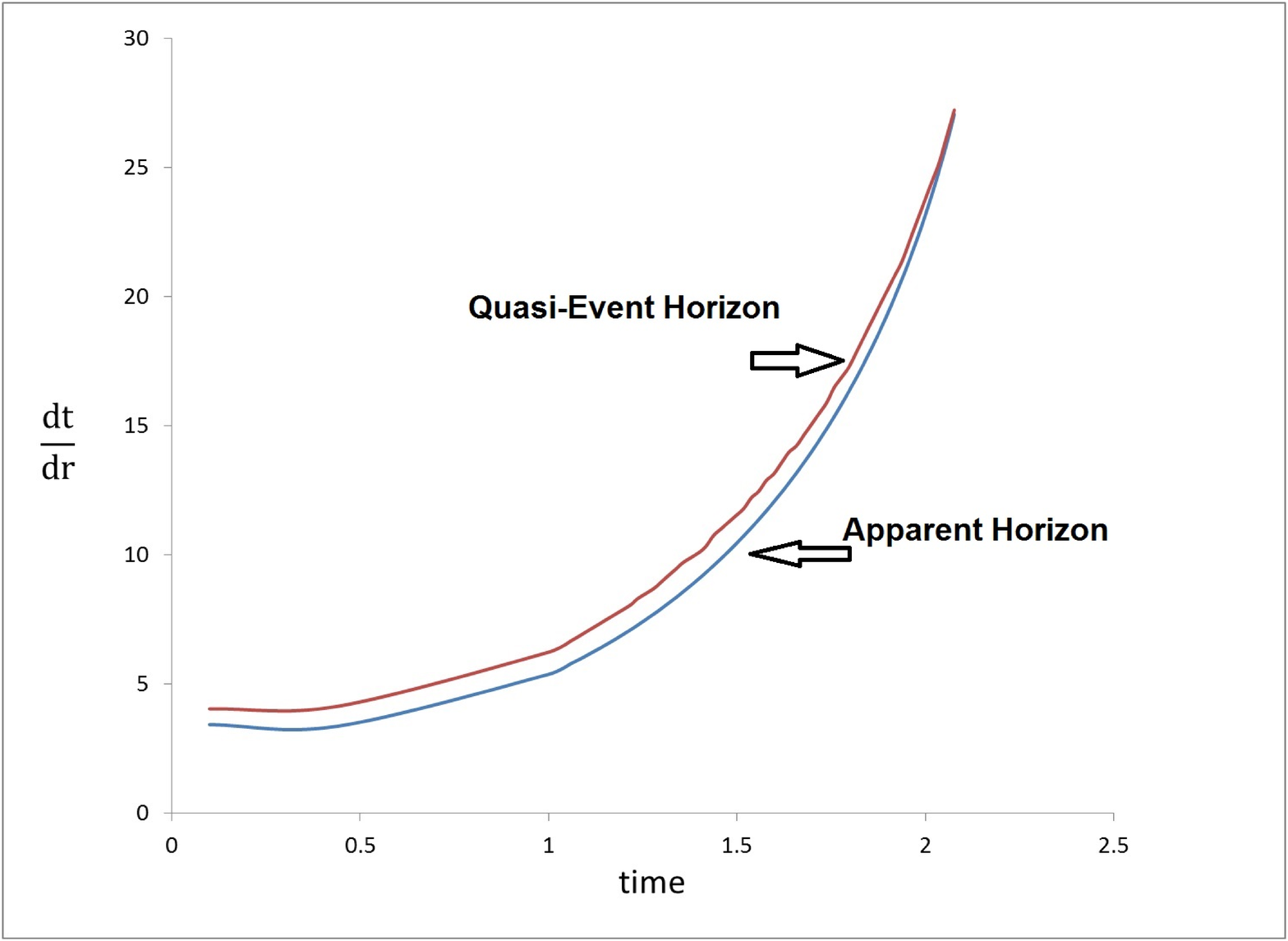}}\quad
    \subfigure[The $p=ws(r)\rho$ case: $ \frac{dt}{dr}|_{AH}< \frac{dt}{dr}|_{null}$  on the apparent horizon.  Qualitatively,  there
is no difference to the Fig.(a).\label{horizon2}]{\includegraphics[width=0.5\linewidth]{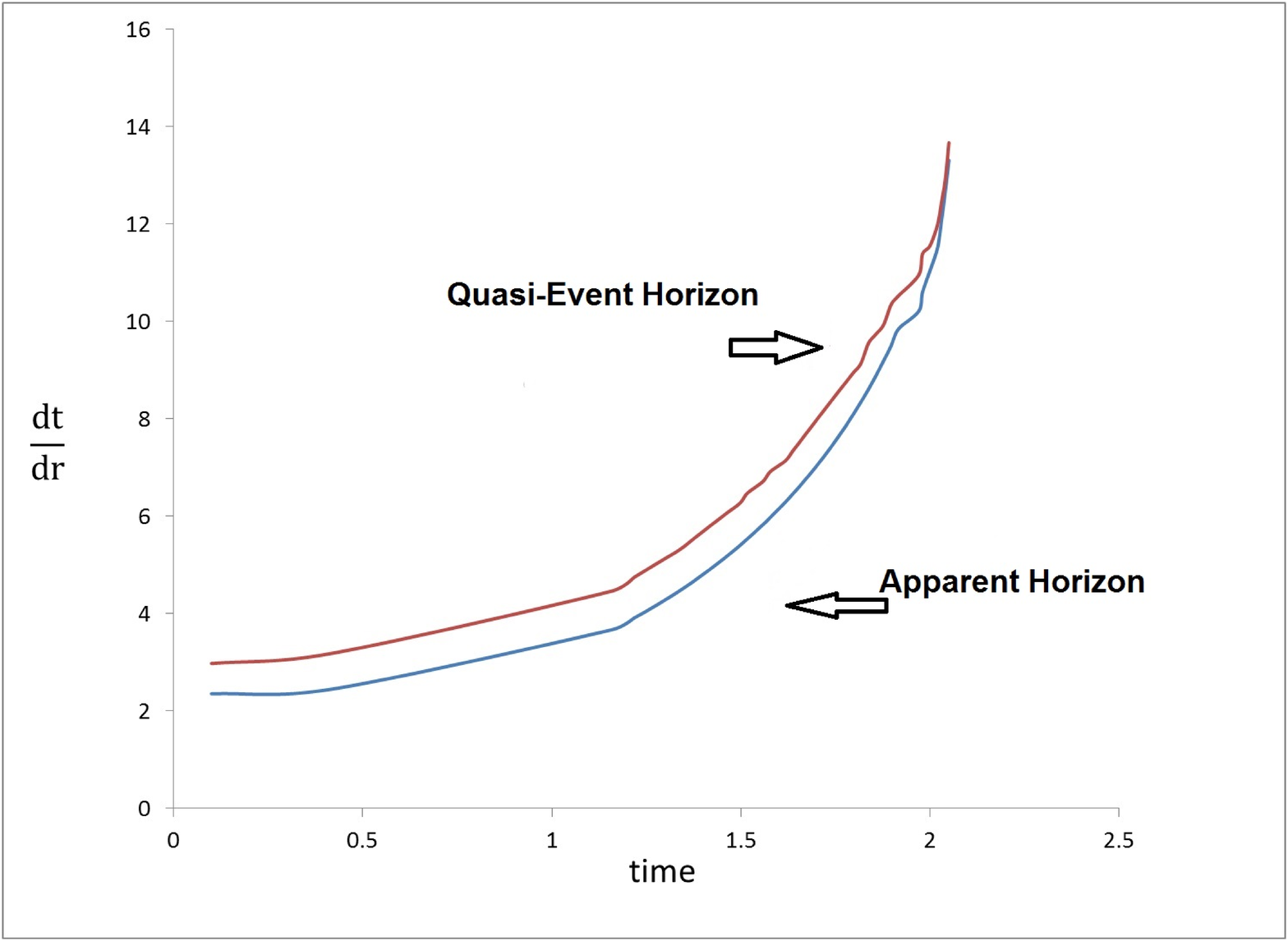}}
  }
  \caption{ $ \frac{dt}{dr}|_{AH}< \frac{dt}{dr}|_{null}$  on the apparent horizon.  Qualitatively,  there
is no difference between two equation of states.}
  \label{horizon3}
\end{figure}

We now show how the dynamical horizon of our cosmological black hole becomes a \emph{slowly evolving horizon} at late times. Let's first define the evolution
parameter $c$ such that the tangent vector to the dynamical horizon, $V $, is given by
\begin{equation}
V^\mu=\ell^\mu-c n^\mu,
\end{equation}
where the two vectors $\ell^a$ and $n^a$ are normal null vectors on a space-like two surface $S$ in $(t,r)$ plane (see \cite{ashtekar02}).
We expect $c$ to go to zero at late times in order for our dynamical horizon to become a \emph{slowly evolving horizon}. In the case of our \L  $~$ model
$c$ is calculated to be
\begin{eqnarray}
c=2\frac{M'+w M'}{M'-w M'-R'}|_{AH}.
\label{black hole}
\end{eqnarray}
 The result of the numerical calculation  for different equation of states and different state functions is given in Fig.(\ref{slowly3}).
The decreasing behavior of the function $c$ in the course of time independent of the equation of state is evident. We may then conclude that the dynamical horizon of the
cosmological black hole tends to a slowly evolving horizon.
\begin{figure}
  \centering
  \mbox{
   \subfigure[The $p=w\rho$ case: the more the pressure the sooner the dynamical horizon becomes a slowly evolving horizon. \label{slowly1}]{\includegraphics[width=0.5\linewidth]{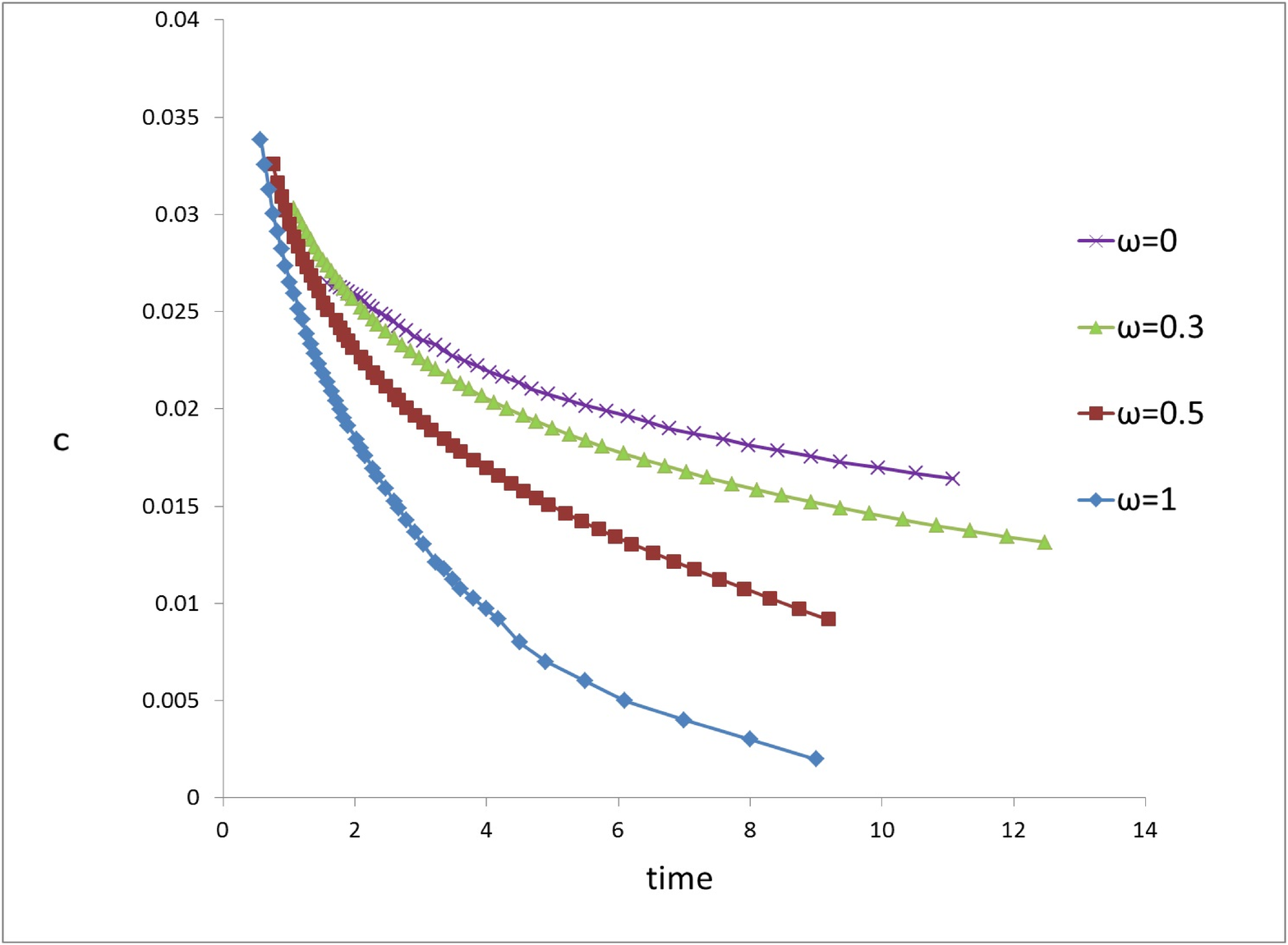}}\quad
    \subfigure[The $p=w\rho s(r)$ case: qualitatively, the same behavior as in Fig.(a).\label{slowly2}]{\includegraphics[width=0.5\linewidth]{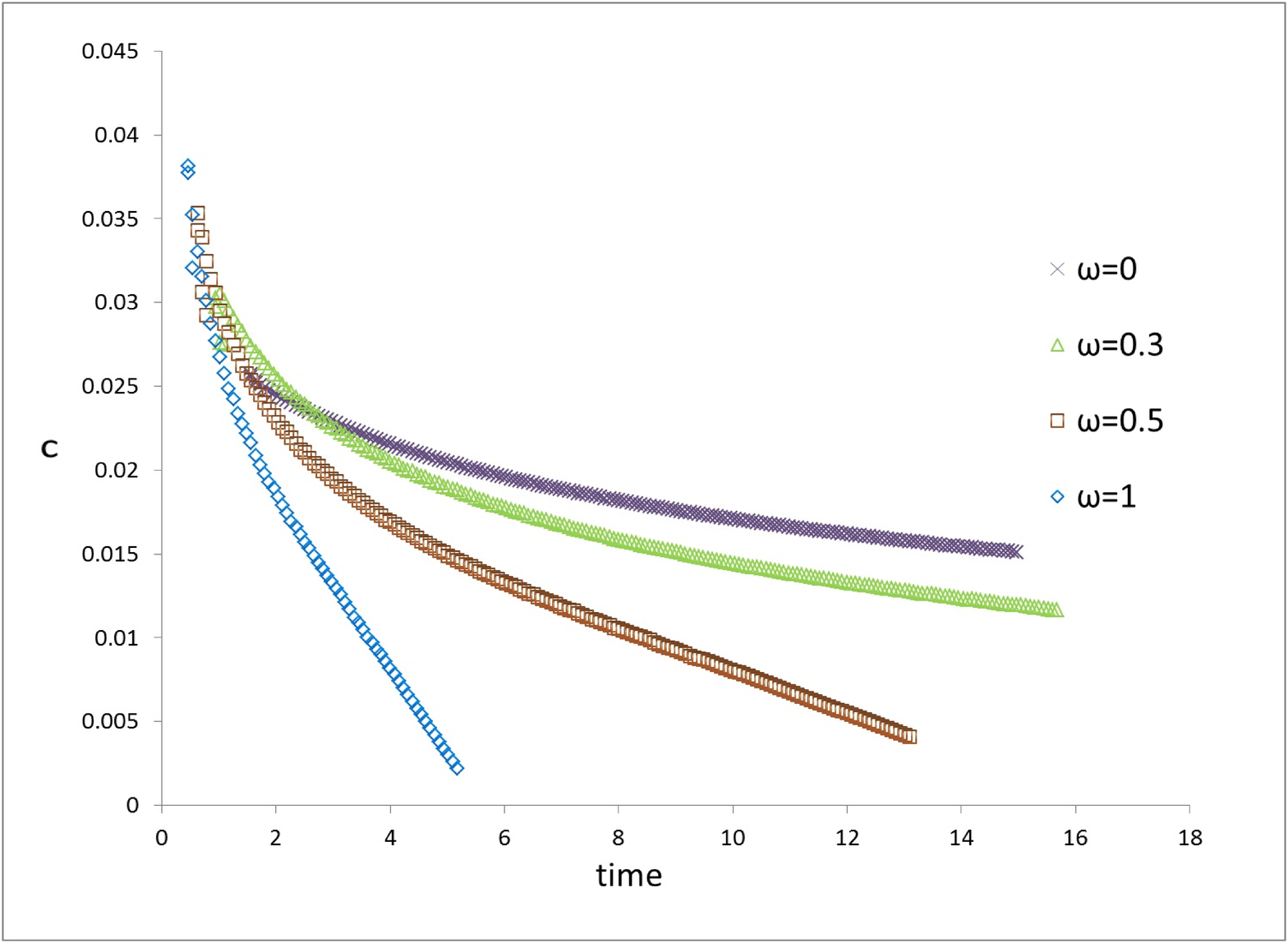}}
  }
  \caption{ Dynamical horizon becomes a slowly evolving horizon.}
  \label{slowly3}
\end{figure}

\subsection{The Effect of  $\Lambda$ on  the Formation of the Apparent Horizon}

 Let us now explore how the addition of a cosmological term  $\Lambda $ to the Einstein equations may influence the formation of the apparent horizon. The location of the apparent horizon  is now defined by  \cite{mangeneral}
 \begin{equation}
  2M=R- \frac{\Lambda}{3}R^3 ~.
      \label{lambda}
 \end{equation}
Now, for a typical structure like a galaxy or a  cluster of galaxies we have $\frac{\Lambda}{3}R^2\ll\frac{2M}{R} $. Therefore, the location of the apparent horizon 
is almost the same as in the case of vanishing cosmological constant, i.e. $2M \approx R $. We may then refer to Fig(\ref{hor}) for the behavior of the apparent horizon.
\begin{figure}
  \centering
  \mbox{
   \subfigure[The $p=w\rho$  case. \label{acc}]{\includegraphics[width=0.5\linewidth]{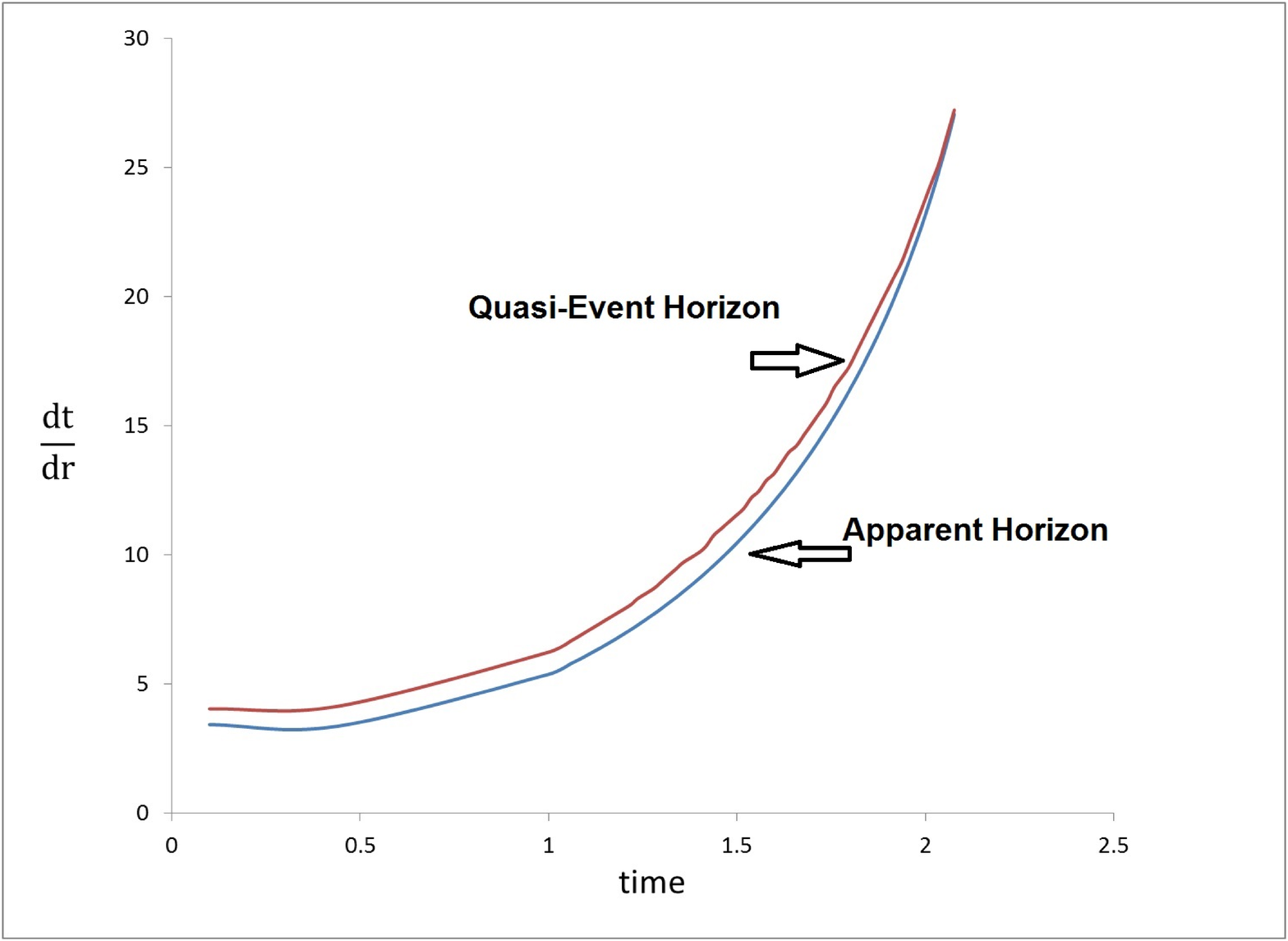}}\quad
    \subfigure[The $p=ws(r)\rho$ case.\label{acc1}]{\includegraphics[width=0.5\linewidth]{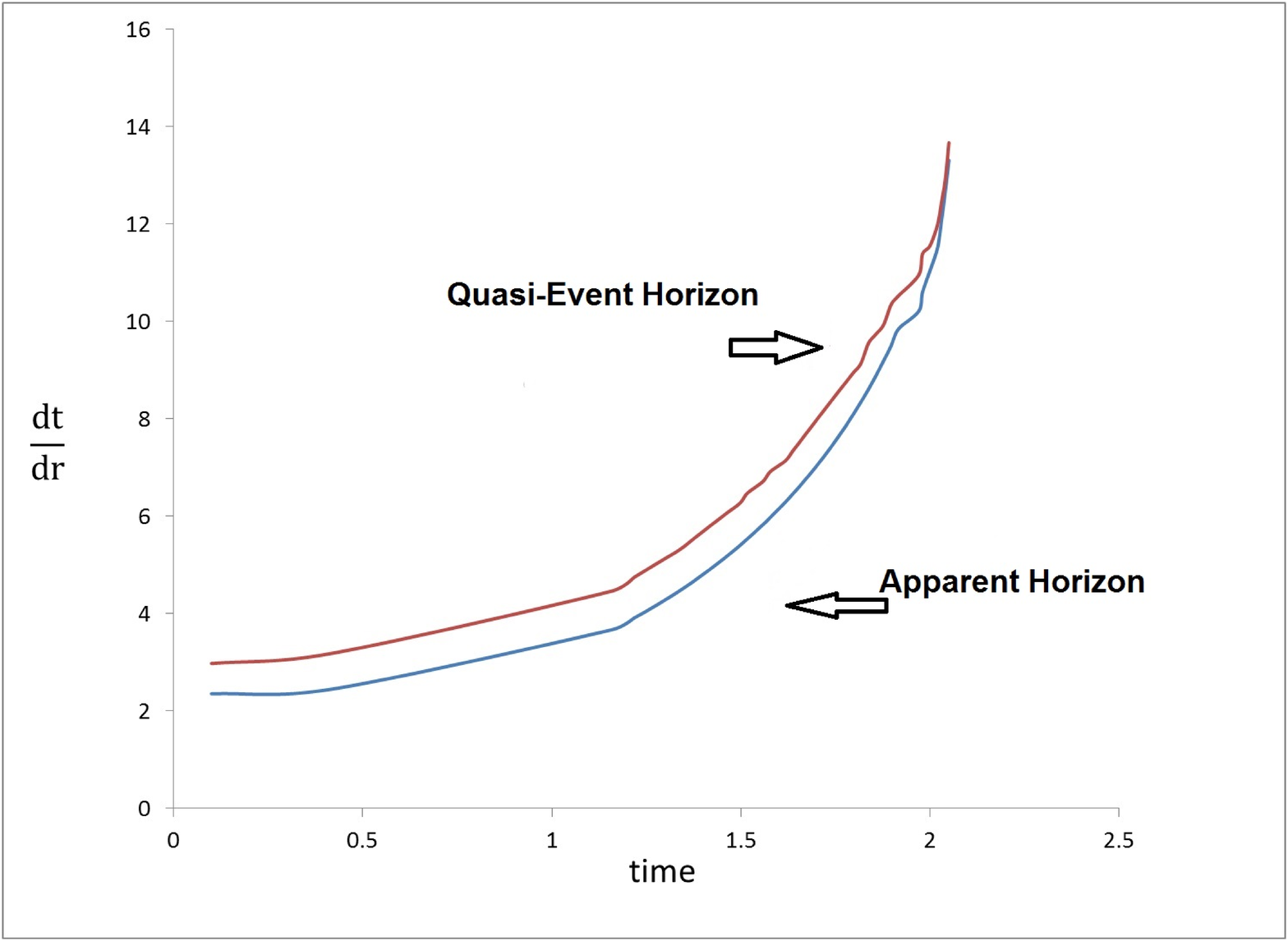}}
  }
  \caption{Adding $\Lambda $ to the Einstein equations does not have any observable effect on the formation of the apparent horizon. }
  \label{hor}
\end{figure}
}

\subsection{ Mass and matter flux}

Due to the expanding background we expect the matter flux into the dynamical black hole to be decreasing and the dynamical horizon to become a
slowly evolving horizon in the course of time\cite{mangeneral}. We know already that there is no unique concept of mass in general relativity corresponding to the
Newtonian concept. The question of what does general relativity tell us about the mass of a cosmological structure in a dynamical
setting was discussed recently \cite{manmass}. It was shown \cite{razbin} that The Misner-Sharp quasi-local mass, $M$, is very close to the
Newtonian mass.

 Let us then take the Misner-Sharp mass for this black hole and calculate the corresponding matter flux into the black
hole. In the case of \L model, the matter flux is given by
\begin{eqnarray}
\frac{dM(r,t)}{dt}|_{AH}=\frac{\partial{M(r,t)}}{  \partial{t}}|_{AH} +\frac{\partial{M(r,t)}}{\partial{r}}\frac{\partial{r}}{\partial{t}}|_{AH}=\dot{M}|_{AH}+M'\frac{\partial{r}}{\partial{t}}|_{AH}
\end{eqnarray}

 The result of the numerical calculation is depicted in Fig.(\ref{flux3}). Note how the pressure decreases the rate of matter flux into the
black hole at the late time.

\begin{figure}
  \centering
  \mbox{
   \subfigure[The $p=w\rho$ case: the rate of matter flux into the black hole decreases with the pressure.  \label{flux1}]{\includegraphics[width=0.5\linewidth]{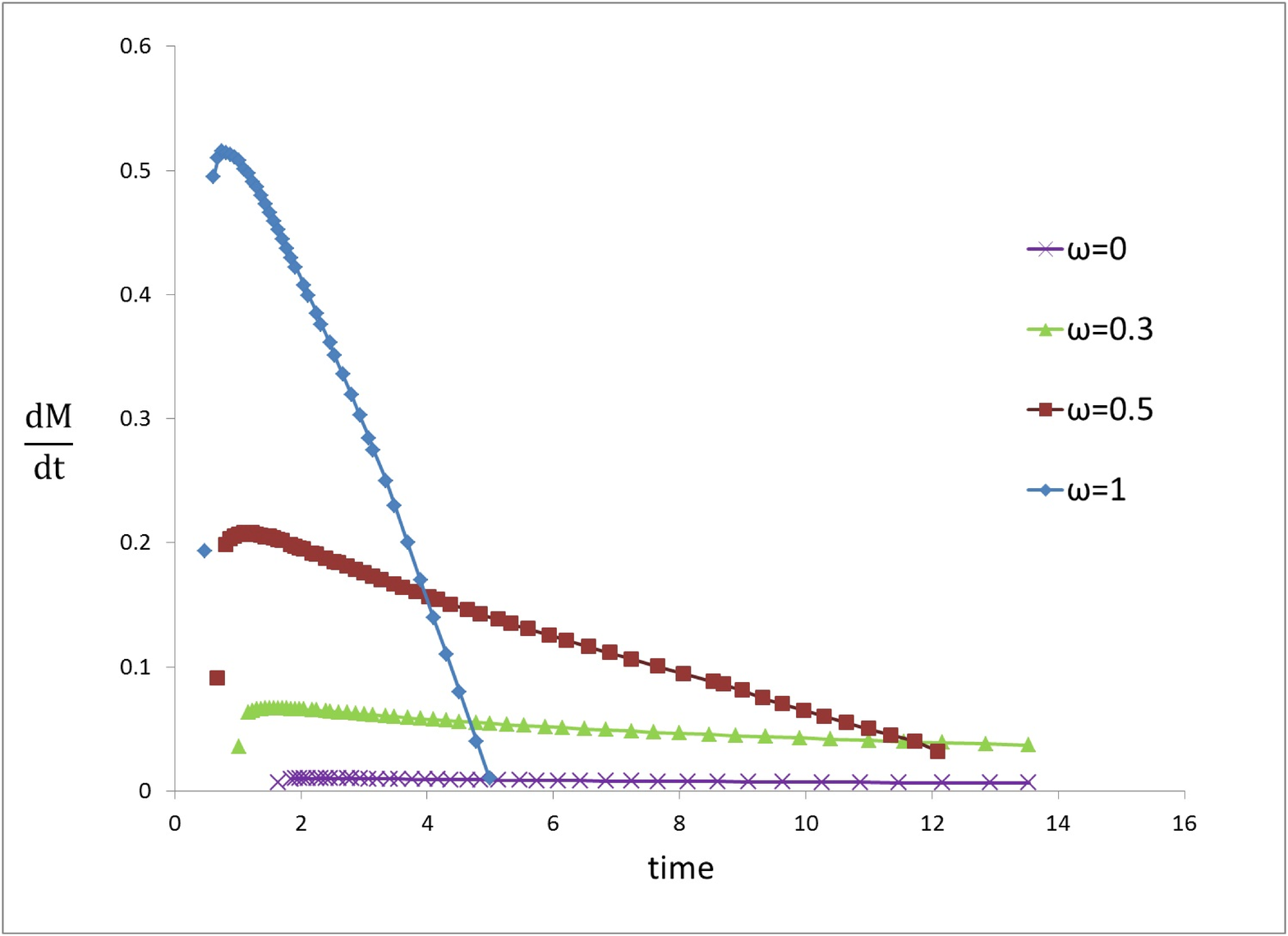}}\quad
    \subfigure[The $p=ws(r)\rho$ case: qualitatively the same behavior as in Fig.(a).\label{flux2}]{\includegraphics[width=0.5\linewidth]{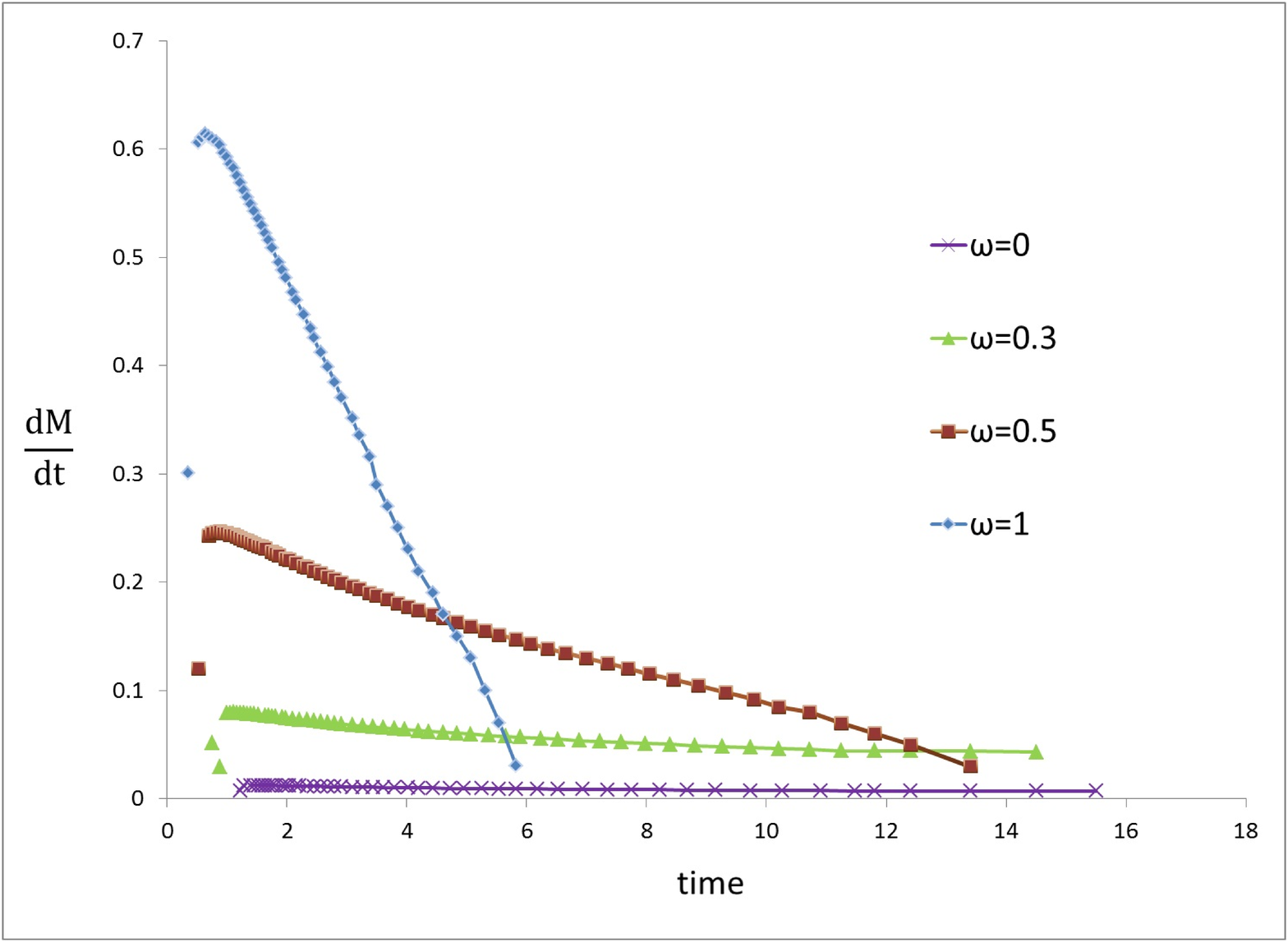}}
  }
  \caption{ The rate of matter flux into the cosmological black hole.}
  \label{flux3}

\end{figure}

\section{Measuring the redshift in the Lema\^{i}tre  metric}

In cosmology we are used to interpret the cosmological redshift according to the homogeneous FRW model. What if the universe is inhomogeneous? In the simplest case we are ready to model a source, a cosmological structure, within an otherwise homogeneous FRW model using our Lema\^{i}tre model. We assume now an observer far from the source of light near the structure. What is then the redshift measured by this observer? Given that the metric is an exact solution of the Einstein equations, we expect the redshift to include all gravitational effects including not only the cosmological redshift but also the gravitational redshift due to the overdensity of the source.    \\ 
The redshift in our model can be obtained as follows. Assume the light ray coming from a source S located near to the structure in the center of our inhomogeneous Lema\^{i}tre model, and the observer O somewhere within the FRW background having the corresponding 4-velocities $u^\mu_{(s)}$ and $u^\mu_{(o)}$. Let $k^\mu\equiv dx^\mu/d\beta$ be the tangent vector to the null geodesic connecting the source to the observer. The corresponding redshift $z$, i.e. frequency shift, is then defined
as\cite{Dwivedi}
\begin{equation}
\label{zed}
1+z={[k_\mu u^\mu_{(s)}]_{P_1}\over [k_\mu u^\mu_{(o)}]_{P_2}},
\end{equation}
where $ [k_\mu u^\mu_{(s)}]_{P_1}$ and\ $  [k_\mu u^\mu_{(o)}]_{P_2}$ are evaluated at the source and observer events. Let us
assume the null geodesics to be radial, i.e. $k^\mu k_\mu=0=k^\mu_{;\nu}k^\nu$. For the metric in equation (1) we then have
\begin{equation}k^r={e^{\sigma}\over e^{\frac{\lambda}{2}}}k^t\Rightarrow {dt\over dr}=
{ e^{\frac{\lambda}{2}}\over e^{\sigma}}.
\end{equation}

Using now the geodesic equation, it is straightforward to show that
\begin{equation}{dk^t\over d\beta}=-[{\sigma'+(-\dot{\sigma}+\dot{\lambda \over 2}){ e^{\frac{\lambda}{2}}\over e^{\sigma}}}]k^tk^r,
\end{equation}
where $\beta$ is an affine parameter. Therefore, \newline{}

\begin{equation}
\label{zed1}
{dk^t\over k^t}=-[{\sigma'+(-\dot{\sigma}+\dot{\lambda \over 2}){ e^{\frac{\lambda}{2}}\over e^{\sigma}}}]d\beta k^r = -[{\sigma'+(-\dot{\sigma}+\dot{\lambda \over 2}){ e^{\frac{\lambda}{2}}\over e^{\sigma}}}]dr.
\end{equation}
Integrating the Eq.(\ref{zed1}) we obtain
\begin{equation}
\label{zed4}
k^t=c_o\exp{(-\int{[{\sigma'+(-\dot{\sigma}+\dot{\lambda \over 2}){ e^{\frac{\lambda}{2}}\over e^{\sigma}}}]dr})},
\end{equation}
 where $c_o$ is a constant.\\
 Now, the 4-velocities are given by
\begin{equation}
\label{zed2}
u^\mu_{(s)}=\delta^\mu_t e^{-\sigma(r_s, t_{s})},\end{equation}
\begin{equation}
\label{zed3}
u^\mu_{(o)}=\delta^\mu_t e^{-\sigma(r_o, t_{o})}.
\end{equation}

Using equations (\ref{zed}), (\ref{zed2}), (\ref{zed3}), and (\ref{zed4}) we obtain the cosmological redshift in the presence of a structure ($z_{CBH}$): 

\begin{equation}
\label{zed55}
1+z_{CBH}=\exp[{(\int_{r_o}^{r_1}[{\sigma'+(-\dot{\sigma}+\dot{\lambda \over 2}){ e^{\frac{\lambda}{2}}\over e^{\sigma}}}]dr)}]\frac{e^{\sigma_{r_o}}}{e^{\sigma_{r_1}}}.
\end{equation}
\begin{equation}
\label{zed5}
1+z_{CBH}=\exp[{(\int_{r_o}^{r_1}[{(-\dot{\sigma}+\dot{\lambda \over 2}){ e^{\frac{\lambda}{2}}\over e^{\sigma}}}]dr)}]=\exp[{(\int_{r_o}^{r_1}[\frac{\partial }{\partial t}{({ e^{\frac{\lambda}{2}}\over e^{\sigma}}})]dr)}].
\end{equation}
Note that there was no need to calculate the $k^r$ due to $k_r u^r=0$.\\
Now, the Eq.(\ref{zed5}) may be integrated numericaly using the equations for $\dot{\sigma}$, $ \sigma' $, and $\dot{\lambda}$ from \cite{mine} for any collapsing structure in an expanding FRW universe. The necessary initial condition may be chosen along the line discussed in \cite{mine}. \\
This redshift includes  the familiar cosmological FRW part, $z_C$, as well as the gravitational redshift, $z_G$ due to the overdensity of the structure (the cosmological black hole). In general, we then expect it to be different from that of the corresponding homogeneous FRW model. In the special case of a homogeneous universe without
a structure it reduces obviously to the familiar FRW cosmological redshift $z_C = \frac{a(t_o)}{a(t_s)}-1$, lacking the contribution from the local gravitational redshift of the overdensity, $z_G$. To see the difference between the exact inhomogeneous
cosmological redshift according to our model and the sum $z_G + z_C$, let's look at some specific models.\\
Take a CBH model with the mass $10^{6} M_{\odot} $ at a distance corresponding to $ z_C = 0.005$ from the observer. The gravitational redshift according to the Schwarzschild metric is given by

\begin{equation}
\label{GR}
z_G\simeq = {1\over (1-\frac{2M(r,t)}{ R_s})}-1.
\end{equation}
Adding to it the FRW cosmological redshift
\begin{equation}
\label{CR}
z_C=\frac{a(t_o)}{a(t_s)}-1,
\end{equation}
should give us the CBH redshift we have already calculated, i.e.

\begin{equation}
\label{CR}
z_{CBH} \simeq z_G+z_C
\end{equation}

 The result of the numerical calculation is given in Fig.(\ref{CBH1}). We see that the exact $z_CBH$ is always larger that the sum of FRW cosmological redshift and the local Schwarzschild gravitational redshift, although the difference is smaller than the observational limit of accuracy. The difference $z_CBH - z_C$ is shown in  Fig.(\ref{zed20}). The difference goes to zero for the source being at distances far from the apparent horizon. There may be, however, cases that this difference is not to be ignored. We leave it to a more detail study in future to see where this difference may be of any cosmological significance.

\begin{figure}[h]
\includegraphics[width = 8cm]{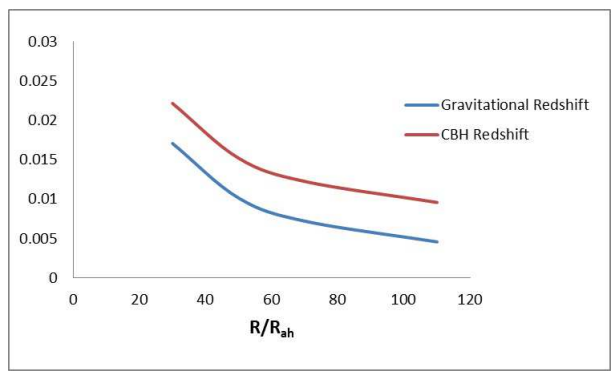}
\caption{ \label{CBH1} Difference of gravitational redshift and CBH redshift for the some source near the CBH ($ R_{ah}=R_{apparent-horizon}$). The observer place is fixed.}

\end{figure}

\begin{table}[h!]
\centering
\begin{tabular}{||c c c c||}
 \hline
$ R/R_{ah}$ & $z_G$ & $z_{CBH}$ & $[z_{CBH} -(z_G+z_C)]/(z_{CBH})$  \\ [1ex]
 \hline\hline
 30 & 0.01709 & 0.022173 & 0.00374 \\
\hline
 50 & 0.01015 & 0.015211 & 0.00401 \\
\hline
 70 & 0.00722 & 0.012263 & 0.00350 \\
\hline
 100& 0.00458 & 0.00959 & 0.00101\\
 [1ex]
 \hline
\end{tabular}
\caption{Numerical values of Fig.(\ref{CBH1})}.
\label{table:1}
\end{table}

\begin{figure}[h]
\includegraphics[width = 8cm]{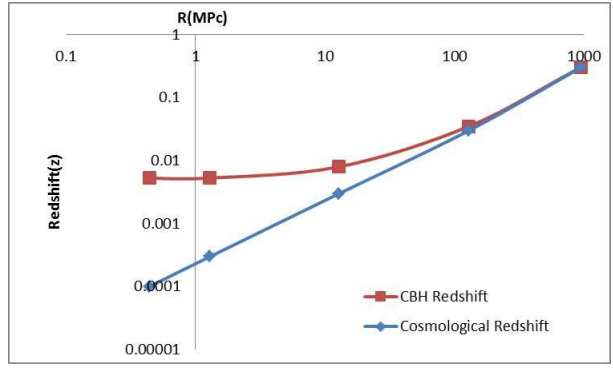}
\caption{ \label{zed20} The difference between cosmological and CBH redshift for several observers located at different physical distances, the source of photon is fixed for all observers.}
\end{figure}

\section{Modeling the evolution of a cluster of galaxies using the Lema\^{i}tre metric: Case of A2061}

Now we are ready to use our exact solution and the corresponding algorithm to study the collapse of a real galaxy cluster within an otherwise
expanding universe using a fluid model. Clusters of galaxies are modeled by a spherically symmetric (isotropic) dark matter halo assumed to dominate the dynamics of the system. The density profile of such a halo is often described by the Navarro-Frenk-White (NFW) profile \cite{NFW}:
\begin{equation}
\label{nfw}
\rho_{NFW}(r)=\frac{\delta_c \rho_c}{\frac{r}{r_s}(1+\frac{r}{r_s})^2},
\end{equation}
where $r_s$ is a scale radius, $\rho_c$ is the critical density, and the characteristic overdensity $\delta_c$ being a function of cluster concentration $c = r_{200}/r_s$ given by
\begin{equation}
\delta_c=\frac{200}{3}\frac{c^3}{\ln(1+c)-\frac{c}{1+c}}.
\end{equation}
The parameter $ r_{200}$ is the radius at which
the average interior density is $200~\rho_c$ being approximately equal to the virialized overdensity, $\rho_{virial}=178~\rho_c$.
The NFW profile in Eq.(\ref{nfw}) has three undesirable features to be applied to our case of a cosmological structure:\newline{} 1) The
density tends to infinity at small r. \newline{}2) The NFW mass diverges at large r and can not be matched to the with FRW mass; \newline{}3) Due to the lack of a void it doesn't represent a density profile for a structure within an expanding universe. \newline{} 
 We have then to modify it to remedy these deficiencies. To resolve the first problem, we introduce a
maximum density at very small r:
\begin{equation}
\label{nfw1}
\rho_{1~NFW}(r)=\frac{\delta_c \rho_c}{(\epsilon_c + \frac{r}{r_s})(1+\frac{r}{r_s})^2}.
\end{equation}
The second problem will be resolved by introducing the truncation radius $r_t$:
\begin{equation}
\label{nfw2}
\rho_{2~NFW}(r)=\frac{\delta_c \rho_c}{(\epsilon_c + \frac{r}{r_s})(1+\frac{r}{r_s})^2}{(\frac{r^2_t}{r^2+r^2_t})^2}.
\end{equation}
In order to resolve the third problem, we add a Gaussian density profile to match the overdensity region through a void to the cosmological background as is necessary for any exact solution of Einstein equations (see \cite{mansouri-khakshournia} and \cite{mojahed}):
\begin{equation}
\rho_G(r) =a \exp(-\frac{(r-r_1)^2}{r_0}).
\end{equation}
The parameters $r_0$ and $r_1$ give us the freedom to adjust the location of the void as desired.
 For our model cluster, A2061 \cite{Yoon}, we choose $ c =3$, $a = 1  $Mpc, and $\epsilon_c = 0.1$ to have $r_{200 }=1.723  $Mpc .
Our metric solution has to approach FRW at large r outside the cluster. Therefore, we add a homogeneous background density such that the final density 
profile at $t_0$ becomes
\begin{equation}
\rho_{cluster}(r) =\rho_{2~NFW}(r)-\rho_G(r) + \rho_c.
\end{equation}
Note that the Gaussian profile must be added in such a way that 
\begin{equation}
\int_{0}^{r_1} r^2 \rho_G dr=\int_{0}^{r_1} r^2 \rho_{2}(r) dr,
\end{equation}
where $r_1$ is the distance at which the density approaches the background critical density. It guaranties that the mass due to the overdensity in the structure is compensated by the underdensity region within the void (See Fig. (\ref{NFW1})).\\
Now, by choosing $R(t_0 , r)=r $ and integrating Eq. (\ref{pressure}) to determine $M(t_0 ,r) $ as well as having the equation of state, we can solve the coupled evolution PDEs numerically. To have zero pressure at FRW background, a good choice for the equation of state is $ p(t,r) = w(\rho(t,r)-\rho_c)$, in which $\rho_c  $ is just a function of time. The other initial functions are as before. Fig.(\ref{NFW1}) shows the evolution of the galaxy cluster. By modifying the NFW density profile to match the exact general relativistic requirements we have therefore arrived at a model cluster based on the exact solution. We may even go further and see how good we may rely on the observation of the red shift of such a cluster and compare it to the model reflected in Eq.(\ref {zed5}).\\
Based on the redshift of clusters estimated with spectroscopic data and the redshift of the brightest cluster galaxies (BCGs) \cite{Yoon} we may take the cosmological redshift of the galaxy cluster A2061 to be $z=0.07845$, being located at a distance around 330 Mpc. We may therefore model our galaxy cluster such that its cosmological redshift is $z_c=0.07845$; The Eq.(\ref{zed5}) will then tell us if we need to add the gravitational redshift as part of the cosmological one.\\
Fixing the observer distance at 330Mpc, and solving the Eq.(\ref{zed5}) we see that the gravitational redshift is about two orders of magnitude smaller than the cosmological redshift. Therefore, we have a well defined dynamical model of the structure including a modified density profile fitted well with observation. The effect of gravitational redshift may be much bigger for supermassive blackholes \cite{Pollack:2014rja}. Any other density profile may also be modified and adapted to requirements as indicated in this section.
\begin{figure}[h]
\includegraphics[width = 8cm]{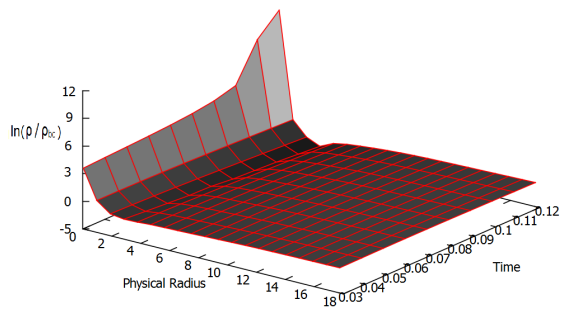}
\caption{ \label{NFW1} Density evolution of the galaxy cluster.}
\end{figure}

\section{DISCUSSION}

We have studied the evolution of a structure made of perfect fluid with non-vanishing pressure as an exact solution of Einstein equations within an otherwise
expanding FRW universe. The structure boundary is separated by a void from the expanding part of the model which is very much like
a FRW universe already near by the void. We have noticed a counter-intuitive pressure effect somewhere inside the structure where the existence
of the pressure slows down the collapse like a classical fluid in contrast to distances far from the structure.
The collapsed region develops to a dynamical black hole with a space-like apparent horizon, in contrast to the Schwarzschild black hole. This apparent
horizon tends to a slowly evolving horizon becoming light-like at late times with a decreasing mater flux into the black hole. We have, therefore, to
conclude that the mere existence of a cosmological matter, even dust, may have significant effect on the central black hole differentiating
it from a Schwarzschild one irrespective of how small the density outside the structure is.\\
The light properties of these cosmological black holes can be interesting because most of the information about the black holes, galaxies and clusters are coming from their lights. We have investigated the redshift of a light emitted near by a cosmological structure to a distant observer. It was shown that the exact CBH redshift of light includes both local gravitational and cosmological redshift of the structure. Therefore, in the era of precision cosmology we may be forced to consider the effects of inhomogeneities as seen in the CBH cosmological model for high redshift surveys.\\
We have also seen how to generalize the existing models of density profile for cold dark matter within large structures using the results of our structure model. Although the gravity may be too weak near a large cosmological structure, we can not use the Newtonian approximation due to the non-local or quasi-local cosmological effects.

\end{document}